\documentclass[epj]{webofc}
\usepackage[varg]{txfonts}   
\usepackage{bm}
\newcommand{\be}{\begin{equation}}
\newcommand{\ee}{\end{equation}}
\newcommand{\bea}{\begin{eqnarray}}
\newcommand{\eea}{\end{eqnarray}}

\newcommand{\h}{\mathfrak h}

\newcommand{\OO}{\mathcal O}

\newcommand{\IR}{{\rm \mbox{{\tiny  IR }}}}
\newcommand{\dil}{{\rm \mbox{{\scriptsize  dil}}}}

\newcommand{\phys}{{\rm \mbox{{\scriptsize  phys}}}}

\wocname{EPJ Web of Conferences}
\woctitle{ICNFP 2015}
\begin{document}
\selectlanguage{english}
\title{On light dilaton extensions of the Standard Model~\thanks{Presented by E.~Meg\'{\i}as at the 4th International Conference on New Frontiers in Physics (ICNFP 2015), 23-30 August 2015, Kolymbari, Crete, Greece.}}
%
%

\author{Eugenio Meg\'{\i}as\inst{1}\fnsep\thanks{\email{emegias@mppmu.mpg.de}} \and
        Oriol Pujol\`as\inst{2}
\and
        Mariano Quir\'os\inst{2,3,4}
}

\institute{Max-Planck-Institut f\"ur Physik (Werner-Heisenberg-Institut), F\"ohringer Ring 6, D-80805, Munich, Germany
\and
           Institut de F\'{\i}sica d'Altes Energies (IFAE), The Barcelona Institute of  Science and Technology (BIST), Campus UAB, E-08193, Bellaterra (Barcelona), Spain       
\and
           Instituci\'o Catalana de Recerca i Estudis  
Avan\c{c}ats (ICREA),  Campus UAB, E-08193, Bellaterra (Barcelona), Spain   
\and
           ICTP South American Institute for Fundamental Research \& Instituto de F\'isica Te\'orica, Universidade Estadual Paulista, S\~ao Paulo, Brazil
}

\abstract{%
We discuss the presence of a light dilaton in Conformal Field Theories deformed by a single scalar operator, in the holographic realization consisting of confining Renormalization Group flows. Then, we apply this formalism to study the extension of the Standard Model with a light dilaton in a 5D warped model. We study the spectrum of scalar and vector perturbations, compare the model predictions with Electroweak Precision Tests and find the corresponding bounds for the lightest modes. Finally, we analyze the possibility that the Higgs resonance found at the LHC be a dilaton.
}
\maketitle

\section{Introduction}
\label{sec:intro}

The discovery of a resonance at the LHC with the properties of a Standard Model-Higgs in 2012 has incited an increasing interest in the study of the nature of the Higgs boson, as this will provide very likely one of the most important indications for the detection of {\it new physics}. An explanation of the light Higgs mass is quite difficult nowadays without the use of unnatural fine tuning of parameters, an issue that is known as the hierarchy problem, see e.g. Ref.~\cite{Giudice:2013nak} and references therein. So, it would be very interesting to study in what kind of scenarios the Standard Model (SM) can be embedded such that the hierarchy problem is solved. One possibility is that the SM be part of a nearly-conformal sector. These models can be realized by means of strong dynamics/extra-dimensions and the Electroweak scale would arise from the spontaneous breaking of conformal invariance (SBCI). It has been recently appreciated that the SBCI can be naturally associated with the appearance of a light dilaton in the spectrum~\cite{CPR,Bellazzini:2013fga,Coradeschi:2013gda,Megias:2014iwa,Cox:2014zea,Megias:2015nya}. Within this realization, the dilaton would be the lightest state of the strong/extra-dimensional sector. Going further, it comes out the intriguing possibility that, in fact, the dilaton can play the role of a {\it Higgs impostor}, so that the $125$ GeV resonance would then be the first instance of new physics. The exploration of this possibility will be one of the main motivations of this work. We will do this by considering a class of soft-wall scenario that allows to fully exploit the modelling capabilities of these theories.

\section{Light dilatons in extra dimensional models}
\label{sec:light_dilatons}

SBCI leads to the existence of a non-zero value for the vacuum expectation value of the dilaton field, which is the Goldstone boson associated with this symmetry. While Conformal Field Theories (CFT) do not naturally exhibit SBCI, it has been recently proposed in Ref.~\cite{CPR} a mechanism that allows for light dilatons in Quantum Field Theories that are close to conformal invariance. This is based on the consideration of certain deformations of CFTs
\begin{equation}
{\cal L} = {\cal L}^{\rm CFT} - \lambda \, {\cal O} \,, \label{eq:L_def}
\end{equation}
where ${\cal O}$ is a nearly marginal operator, i.e. ${\rm dim}({\cal O}) = 4-\Delta$ with $\Delta \ll 1$. A holographic realization of this mechanism has been checked to work in Refs.~\cite{Bellazzini:2013fga,Coradeschi:2013gda,Megias:2014iwa,Cox:2014zea} (see also~Ref.~\cite{Megias:2015nya}). In this Section we describe this mechanism in generic 5D warped models, and then particularize the results to a specific benchmark model.

\subsection{Renormalization Group flows and CFT deformations}
\label{subsec:RG}

\begin{figure*}[htb]
\includegraphics[width=6.7cm,clip]{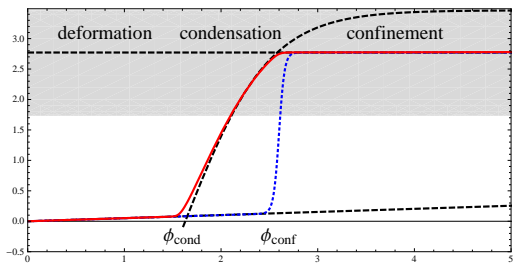}  \hspace{2pc}%
\begin{minipage}{14.8pc}
\vspace{-2cm}\caption{$3\frac{V^\prime(\phi)}{V(\phi)}$ (dotted blue) and $-\beta(\phi)$ (solid red) as a function of $\phi$. The three regimes (deformation, condensation and confinement) are clearly distinguishable in $\beta(\phi)$. We have con\-si\-de\-red Model A of Ref.~\cite{Megias:2014iwa}, which is characterized by $\langle \OO \rangle_\phys \ne 0$.}
\label{fig-1}      
\end{minipage}
\end{figure*}

 Let us consider a coupled scalar-gravitational system in 5D defined by the action
\begin{equation}
S = M^3\int d^4xdy\sqrt{-g}\left(R-\frac{1}{2}(\partial_M \phi)^2
-V(\phi)\right) -M^3 \sum_{\alpha}\int d^4x dy \sqrt{-g}\,2\mathcal V^\alpha(\phi)\delta(y-y_\alpha)  \,, \label{eq:S}
\end{equation}
where $\phi$ is a scalar field, ${\mathcal V}^\alpha \; (\alpha=0,1)$ are UV and IR 4D brane potentials located at $y_0 =y(\phi_0)=0$ and $y_1=y(\phi_1)$ respectively,  and $M$ is the 5D Planck scale. The deformed CFT of Eq.~(\ref{eq:L_def}) can be realized by a {\it domain wall} geometry in which the scalar field develops a profile in the extra dimension~$y$ of the form 
\begin{equation}
ds^2 = dy^2 + e^{-2 A(y)} \eta_{\mu\nu} dx^\mu dx^\nu \,, \qquad \phi = \phi(y) \,. \label{eq:metric}
\end{equation}
By seeking for solutions of the equations of motion with the near boundary expansion
\begin{equation}
\phi(y) = \lambda \; e^{\Delta y} + \langle \OO \rangle \; e^{(4-\Delta) y} + \cdots  \,, \qquad y \to -\infty \,,
\end{equation}
one can identify the deformation parameter and the condensate as the leading and sub-leading mode respectively. The holographic Renormalization Group flows can be studied better with the $\beta$-function, defined holographically as $\beta(\phi) := - \frac{\partial \phi}{\partial A}$. The equations of motion lead to the following first order differential equation
\begin{equation}
\beta(\phi)\,\beta^\prime(\phi) = \frac{1}{2}\left(\beta(\phi)^2-24\right) \left(\frac{\beta(\phi)}{3} + \frac{V^\prime(\phi)}{V(\phi)}\right) \,. \label{eq:beta_eq}
\end{equation}
Associated with this equation there is an integration constant which is identified as the condensate~$\langle \OO \rangle$ of the operator along the deformation direction, Eq.~(\ref{eq:L_def}), and the boundary condition that fixes the physical value of the condensate $\langle \OO \rangle_\phys$ is that the IR end of the flow be the least singular possible, see e.g.~\cite{Papadimitriou:2007sj,Megias:2014iwa,Megias:2015nya}. We are interested in {\it confining flows}, i.e. those which lead to a discrete spectrum in all sectors. In the absence of the IR brane, these flows are characterized by the IR limit 
\begin{equation}
\beta(\phi) \to \beta_\infty  \qquad \textrm{with} \qquad \sqrt{6} < -\beta_\infty < 2\sqrt{6} \,,
\end{equation}
where the lower and upper bounds correspond to the confinement and good IR singularity criterion respectively. A typical profile for such flows is displayed in Fig.~\ref{fig-1}. It is interesting to see that the flow has three different regimes: i) deformation-dominated regime in the UV, ii) condensate-dominated regime, and iii) confinement region in the IR.

\subsection{Spectrum of excitations and type of dilatons}
\label{subsec:spectrum}

To compute the scalar spectrum and in particular its lightest mode, known as the radion/dilaton, we have to consider a scalar perturbation of the background, and this leads to a contribution in the metric of the form~$ds^2 \sim e^{-2(A+F)}$. Then the equation of motion of the excitation modes $F_n$ is written as~\cite{Csaki:2000zn}~\footnote{The dots stand for derivatives with respect to $y$.}
\begin{equation}
\ddot F_n - 2\dot A \dot F_n - 4\ddot A F_n - 2 \frac{\ddot \phi}{\dot \phi} \dot F_n + 4\dot A \frac{\ddot \phi}{\dot \phi} F_n =  - e^{2A} m_n^2 F_n \,. \label{eq:eomscalar}
\end{equation}
This is supplemented with boundary conditions on the branes  $\dot F_n |_{y_\alpha} = f({\mathcal V^\alpha}, m_n)$, which are obtained by integrating in a small interval around $y=y_\alpha$. An approximate solution of Eq.~(\ref{eq:eomscalar}) assuming light modes, i.e. $m_n^2 \ll \Lambda_\IR^2$ where $\Lambda_\IR$ is the mass gap in the spectrum, leads to a {\it mass formula} which allows to distinguish between two kind of light dilatons~\cite{Megias:2014iwa,Megias:2015ory}:
\begin{itemize}
\item {\it Hard dilatons}: those dominated by the value of the $\beta$ function at the IR brane location. The mass of the dilaton scales like $m_\dil^2 \sim \beta_\IR^2 \Lambda_\IR^2$. This is the realization that has been discussed in Refs.~\cite{Bellazzini:2013fga,Coradeschi:2013gda}.
\item {\it Soft dilatons}: those dominated by the value of the $\beta$ function in the condensation scale. In this case a light dilaton can be realized whenever i) the $\beta$ function is small at the condensation scale $(\beta_{cond} \ll 1)$ and ii) the rise towards confinement $(\beta_{conf} \sim 1)$ is fast enough. Then 
\begin{equation}
m_\dil^2 \sim (\beta_{cond})^{\kappa} \Lambda_\IR^2  \; \ll  \; \Lambda_\IR^2   \,,
\end{equation}
where the precise value of the constant $\kappa > 0$ depends on the particular model~\cite{Megias:2014iwa,Megias:2015nya}. The dilaton turns out to be naturally light for nearly marginal deformations, for which $\; \beta_{cond} \sim \Delta \ll \beta_{conf}$.
\end{itemize}

The dilaton corresponds to the fluctuation of the condensate, regardless of whether $\langle \OO \rangle$ vanishes or not. In the following we will simplify the model by setting $\langle \OO \rangle = 0$, as the physics is qualitatively similar to the one from models with $\langle \OO \rangle \ne 0$. This can be done by just considering an analytic $\beta$ function, as then the integration constant in Eq.~(\ref{eq:beta_eq}) is set to vanish. We choose 
\begin{equation}
\beta(\phi) = -6ac\left[1+e^{-a\phi}  \right]^{-1} \,,
\end{equation}
where $c$ and $a$ are parameters which govern the IR value of $\beta$ and its slope at $\phi=0$ respectively. We assume that the brane potentials dynamics have fixed $(\phi_0, \phi_1)$ to solve the hierarchy problem, i.e. $A(\phi_1) \simeq 35$, where we have normalized the warp factor to $A(\phi_0)=0$.

\begin{figure*}[htb]
\begin{tabular}{cc}
\includegraphics[width=6.65cm,clip]{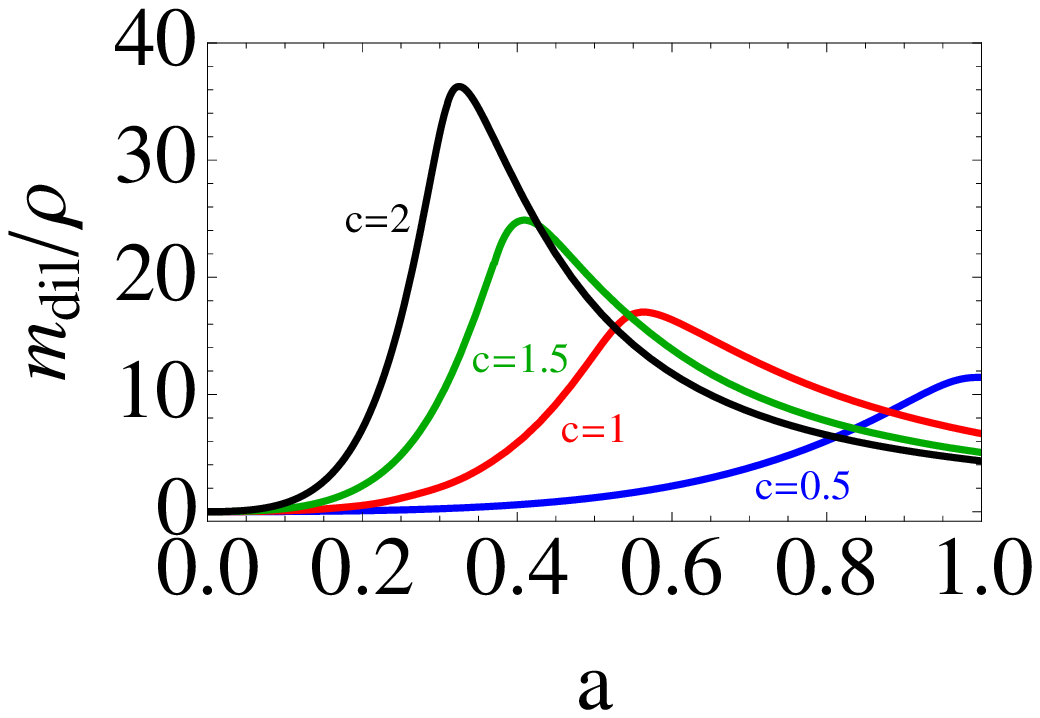} &
\includegraphics[width=6.65cm,clip]{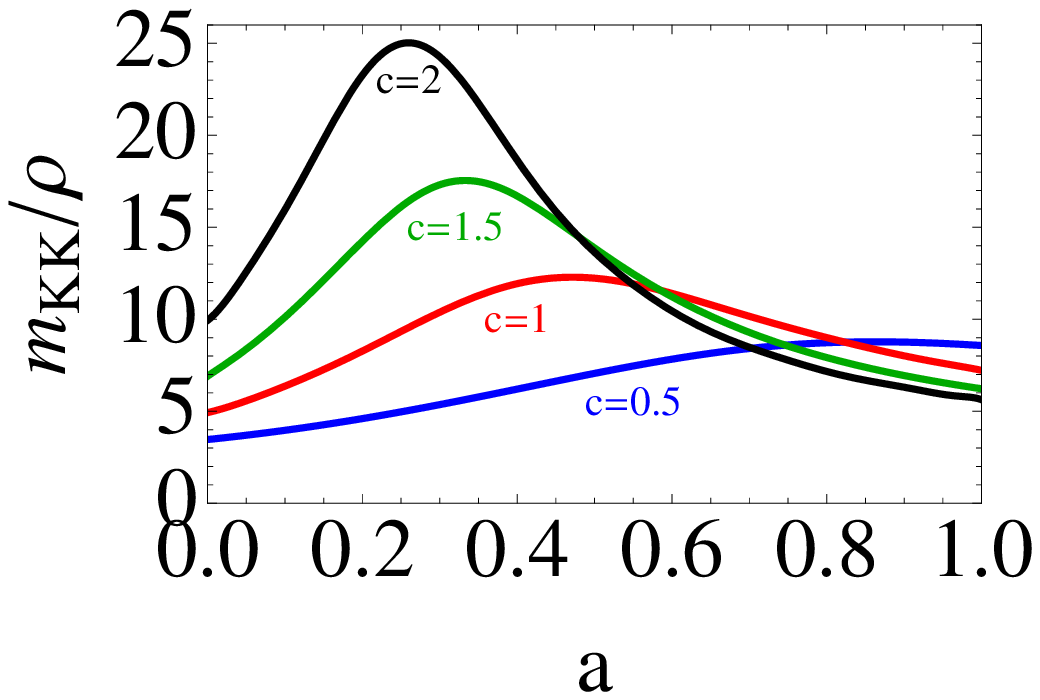}  \\
(A)
     &     
         (B)
\end{tabular}
\caption{(A) Mass of the dilaton as a function of the parameter~$a$, normalized to $\rho = k \, e^{-A(y_1)}$. (B) KK gauge boson masses as a function of the parameter $a$.  In these figures we have plotted the results for $c=0.5, 1, 1.5$ and $2$, and have considered $\phi_1=5$.}
\label{fig-2}
\end{figure*}

We show in Fig.~\ref{fig-2} the dilaton mass obtained from a numerical solution of Eq.~(\ref{eq:eomscalar}). The realization of the dilaton for $a \lesssim 0.6/c$  and $a \gtrsim 0.6/c$ is hard and soft respectively, and the existence of the peak is a signal of the change of regime between both pictures. We also display in Fig.~\ref{fig-2} (B) the spectrum of vector perturbations, which we refer in the following as KK gauge bosons. See Ref.~\cite{Megias:2015ory} for details.

\section{Electroweak breaking}
\label{sec:EW}

\begin{figure*}[htb]
\begin{tabular}{cc}
\includegraphics[width=6.65cm,clip]{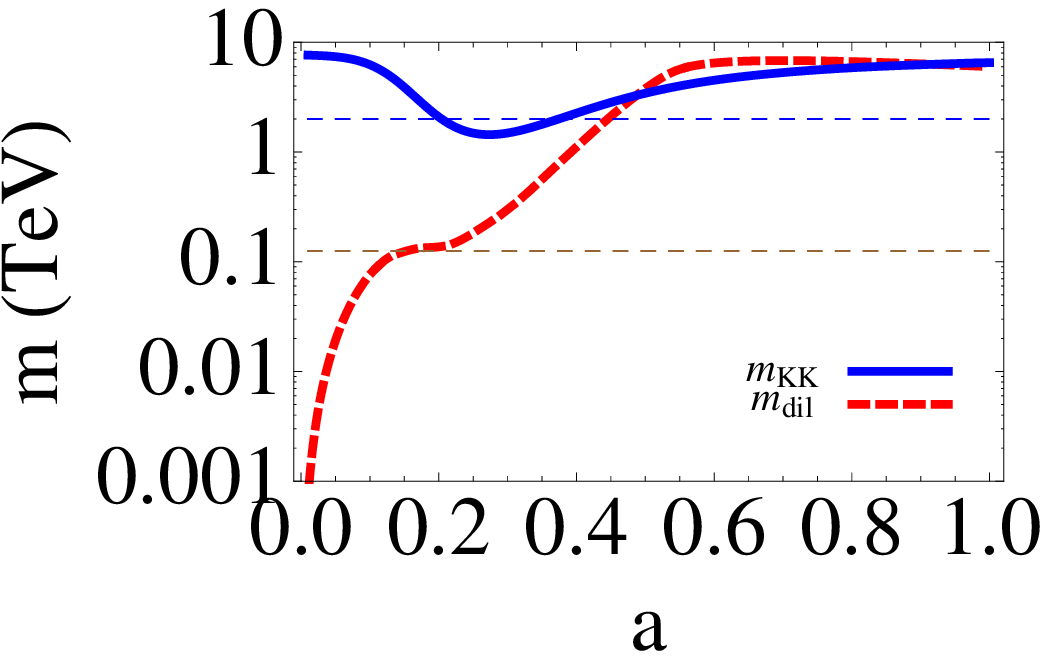} &
\includegraphics[width=6.65cm,height=4.3cm,clip]{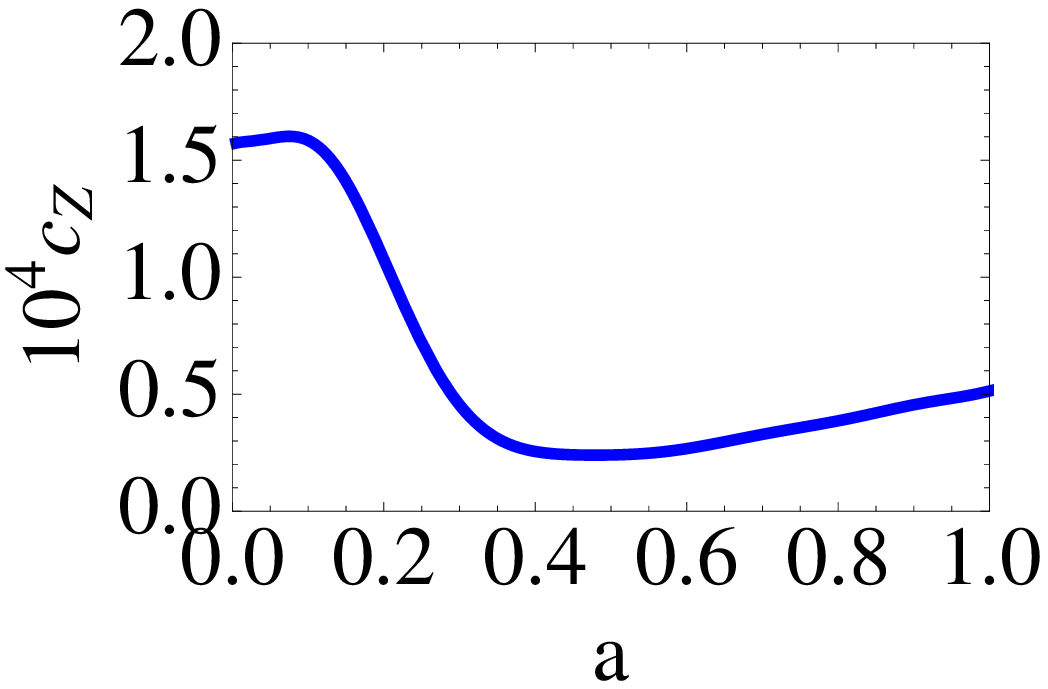}  \\
(A)
     &     
         (B)
\end{tabular}
\caption{(A) Bound on KK mass (solid line) as function of $a$ from electroweak observables. The corresponding dilaton mass is in dashed line. We have drawn horizontal dashed lines corresponding to 125 GeV and 2 TeV. (B)~Radion coupling to the $Z$ boson as a function of $a$. The result for W bosons is $c_W = (m_W/m_Z)^2 \cdot c_Z  \approx 0.78 \, c_Z$. In these figures we have considered $c=1$ and $\phi_1 = 5$.
}
\label{fig:KKdil}
\end{figure*}

We will now introduce the electroweak sector in the theory. Let us consider the SM propagating in the 5D space described in Section~\ref{sec:light_dilatons}. In addition to the 5D ${\rm SU}(2)_L\times {\rm U}(1)_Y$ gauge bosons $W^i_M(x,y)$, $B_M(x,y)$ with $i=1,2,3$ and $M=(\mu,5)$, we define the SM Higgs as
\begin{equation}
H(x,y)=\frac{1}{\sqrt 2}e^{i \chi(x,y)} \left(\begin{array}{c}0\\h(y)+\xi(x,y)
\end{array}\right)
\ ,
\label{Higgs}
\end{equation} 
where $h(y)$ is the Higgs background. The action of the model is
\begin{equation}
S_5=\int d^4x dy\sqrt{-g}\left(-\frac{1}{4} \vec W^{2}_{MN}-\frac{1}{4}B_{MN}^2-|D_M H|^2-V(H)\right) \,,
\label{5Daction}
\end{equation}
where $V(H)$ is the 5D Higgs potential. Electroweak symmetry breaking is triggered on the IR brane. After the breaking, the lightest modes are separated by a gap from the KK spectrum, and the masses for the $W$ and $Z$ bosons are then approximately given by~\footnote{See Ref.~\cite{Cabrer:2011fb} for a wide description of the formalism of electroweak breaking by the bulk Higgs.}
\begin{equation}
m_V^2 \approx \frac{1}{y_1} \int_0^{y_1} dy M_V^2(y) \,, \qquad V = W,\, Z \,, \label{eq:mV}
\end{equation}
where the $y$-dependent bulk masses are defined as
\begin{equation}
M_W(y) = \frac{g_5}{2} h(y) e^{-A(y)} \,, \qquad M_Z(y) = \frac{1}{c_W} M_W(y) \,, \label{eq:MV}
\end{equation}
and $g_5$ is the 5D gauge coupling, which is related to the 4D one $g$ by $g_5 = g \sqrt{y_1}$. After solving the equations of motion for the Higgs field $h(y)$, one can obtain from Eqs.~(\ref{eq:mV})-(\ref{eq:MV}) the masses for the gauge bosons. To compare the model predictions with Electroweak Precision Tests (EWPT) a convenient parameterization is using the $(S,T,U)$ variables in~\cite{Peskin:1991sw,Cabrer:2011fb}. From the fitted values for $S$ and $T$ as~\cite{Agashe:2014kda}:
\begin{equation}
T=0.05\pm 0.07,\quad S=0.00\pm 0.08  \,,  \qquad  \textrm{(90\% correlation)}  \,,
\end{equation}
one gets the result in Fig.~\ref{fig:KKdil} (A).
We find a region in $a$ such that $m_{KK}=\mathcal O$(TeV) and $m_\dil \lesssim\mathcal O$(100)~GeV for $c=1$.~\footnote{Extra-dimensional realizations of Higgs impostors require that the KK scale $m_{KK}$ must not exceed a few TeV, because $m_{KK}$ and the electroweak scale $v=246$ GeV are linked by $m_{KK} \lesssim 4\pi v$.} Other values of $c$ lead to similar conclusions.

\section{Coupling of the radion/dilaton to Standard Model matter fields}
\label{sec:coupling}

A light dilaton has similar interactions with matter as the Higgs, so an obvious question arises: Is the Higgs-like resonance found at LHC actually a dilaton? There are previous studies in the literature on the coupling of the radion to SM matter fields, and they usually consider that the matter is localized on the IR brane~\cite{Csaki:2000zn,Csaki:2007ns}. In this work and motivated by EWPT we are assuming that the matter and Higgs fields are localized in the bulk. If the scalar fluctuation decomposes as~$F(x,y) = F(y) {\cal R}(x)$, we can compute the coupling of the dilaton to the massive gauge fields ($W_\mu$ and $Z_\mu$) normalized as
\begin{equation}
\mathcal L_{rad} = - \frac{r(x)}{v}\left\{ 2 c_W\,m_W^2 W_\mu W^\mu+c_Z\,m_Z^2 Z_\mu Z^\mu \right\} \,, \label{eq:LV}
\end{equation}
where $r(x)$ is the canonically normalized radion field with kinetic term $\frac{1}{2}\left( \partial_\mu r \right)^2$. The case $c_W=c_Z=1$ corresponds to the SM Higgs coupling. After expanding Eq.~(\ref{5Daction}) to linear order in the perturbations, and using the massless radion approximation $F = e^{2A}$, one gets the following result for $c_V$  ($V=W,\, Z$)
\begin{equation}
c_V=2\frac{v\, m_V^2}{\sqrt{6} e^{-A_1}M_{Pl} \, \rho^2}\left(\frac{\int e^{-2A}}{\int e^{2A-2A_1}}  \right)^{1/2} \! k y_1 \!\! \int_0^{k y_1}e^{4A-4A_1}\left(  \frac{\int_{0}^{y} h^2 e^{-2A}}{\int_{0}^{y_1} h^2 e^{-2A}}  -\frac{k y}{k y_1}  \right)^2 d(ky)  \,,
\end{equation}
where $M_{Pl}=2.4\times 10^{18}$ GeV is the 4D (reduced) Planck mass. We show in Fig.~\ref{fig:KKdil} (B) the numerical result of $c_Z$ in the same regime of parameters as in Fig.~\ref{fig:KKdil} (A). The values are very small so that  the present dilaton extension of the SM leads to a LHC phenomenology which deviates from SM predictions by an $\mathcal O(10^{-4})$ effect. Unfortunately these tiny effects would be unobservable at the LHC. For the same reason, the possibility of a Higgs impostor is excluded for the present model.

\section{Conclusions and discussion}
\label{sec:conclusions}

We have studied a mechanism in holography that allows for a naturally light dilaton in CFTs deformed by a single scalar operator. Two different realizations of a light dilaton have been identified: i) the {\it hard} realization is induced by the existence of an IR brane, and the dilaton is incarnated by the IR brane location, and ii) the {\it soft} realization, in which the dilaton is controlled by the condensation threshold of the CFT operator. These results have been confirmed in a 5D warped model which is large enough to exploit the full capabilities of the extra-dimensional models.

The extension of the SM with a light dilaton in the 5D warped model leads to dilaton masses that can naturally be of the order of magnitude of the Higgs mass, and KK vector masses of the order of TeV, all of them compatible with Electroweak Precision Tests. However a first study of the coupling to massive gauge fields suggests that the dilaton couplings to SM particles (affecting e.g.~the unitarization of the $V_LV_L$ elastic and inelastic scattering as well as the strength of production of gauge bosons and fermions) are several orders of magnitude smaller than those predicted for the Higgs in the SM in a way that is unobservable at the LHC. For the same reason our results suggest that in the models discussed here the dilaton cannot be a Higgs impostor. In spite of this, some modifications of the warped model that would allow for a sizeable coupling deserve to be studied~\cite{Megias:2015ory}.

\begin{acknowledgement}
The work of O.P. and M.Q.~is partly supported by the Spanish Consolider-Ingenio 2010 Programme CPAN (CSD2007-00042), by MINECO under Grants CICYT-FEDER-FPA2011-25948 and CICYT-FEDER-FPA2014-55613-P, by the Severo Ochoa Excellence Program of MINECO under the grant SO-2012-0234 and by Secretaria d'Universitats i Recerca del Departament d'Economia i Coneixement de la Generalitat de Catalunya under Grant 2014 SGR 1450. The work of M.Q. is also partly supported by  CNPq PVE fellowship project 405559/2013-5. E.M. would like to thank the Institut de F\'{\i}sica d'Altes Energies (IFAE), Barcelona, Spain, the Instituto de F\'{\i}sica Te\'orica IFT UAM/CSIC, Madrid, Spain, and the Departamento de F\'{\i}sica At\'omica, Molecular y Nuclear of the Universidad de Granada, Spain, for their hospitality during the completion of the final stages of this work. The research of E.M. is supported by the European Union under a Marie Curie Intra-European Fellowship (FP7-PEOPLE-2013-IEF) with project number PIEF-GA-2013-623006. 
\end{acknowledgement}


%
%
%


\end{document}